\documentclass[11pt]{amsart}
\usepackage{amsaddr}
\usepackage{geometry}                
\usepackage{graphicx}
\usepackage{amssymb}
\usepackage{epstopdf}
\usepackage{amssymb}
\usepackage{amsmath}
\usepackage{blkarray} 
\usepackage{comment}
\usepackage{latexsym}
\usepackage{amsthm} 
\usepackage{eucal} 
\usepackage{eufrak}
\usepackage{float}
\usepackage[rightcaption]{sidecap}
\usepackage{graphicx}
\usepackage[utf8]{inputenc}
\usepackage[main=english, ukrainian]{babel}
\usepackage{xcolor}
\usepackage{mathrsfs}
\usepackage{array}
\usepackage{framed}
\usepackage[all]{xy}
\usepackage{sidecap}
\usepackage{wrapfig}
\usepackage{multicol}
\usepackage{lscape}
\usepackage{MnSymbol}
\usepackage{mathtools}
\usepackage{cite}
\usepackage{appendix}
\usepackage{chngcntr}
\usepackage{etoolbox}
\usepackage{lipsum}
\usepackage{rotating}
\usepackage[thinlines]{easytable}
\usepackage{hyperref}
\usepackage{empheq}
\usepackage{enumitem}
\usepackage{lineno}
\DeclareMathAlphabet{\mathpzc}{OT1}{pzc}{m}{it}

\theoremstyle{plain} 
\newtheorem{thm}{Theorem}[section] 
\newtheorem{cor}[thm]{Corollary}

\theoremstyle{definition} 
\newtheorem{defn}{Definition}

\theoremstyle{remark} 

\newtheorem{rmk}{Remark}

\title{Sensitivity of metabolic networks:\\ 
perturbing metabolite concentrations}
\author{Nicola Vassena}
\address{Freie Universit{\"a}t Berlin}
\email{nicola.vassena@fu-berlin.de}
\date{\today}

\begin{document}

\providecommand{\keywords}[1]
{
  \small	
  \textbf{\textit{Keywords---}} #1
}

\maketitle

\tableofcontents

\begin{abstract}
Sensitivity studies the network response to perturbations. We consider local perturbations of the concentrations of metabolites at an equilibrium. We investigate the responses in the network, both of the metabolite concentrations and of the reaction fluxes. Our approach is purely qualitative, rather than quantitative. In fact, our analysis is based, solely, on the stoichiometry of the reaction network and we do not require any quantitative information on the reaction rates. Instead, the description is done only in algebraic terms, and the only data required is the network structure. Biological applications include metabolic control and network reconstruction.\\

\smallskip

\end{abstract}

\section{Introduction} \label{intro}

Sensitivity analysis investigates how a network responds to perturbations. Here, we are interested in perturbations, which do not alter the structure of the network. We perturb a network component, for example a metabolite concentration, and we ask which other components have been \emph{influenced} by our direct perturbation, and which have remained unaffected. This concept applies to a large variety of phenomena. Parameter fluctuations, for instance, can easily be induced both by environmental as well as genetic agents. The impact of parameter fluctuations on the network is, therefore, a matter of crucial importance.\\

More specifically, we focus on networks at an \emph{equilibrium}. Here, equilibrium means, simply, that metabolite concentrations are stationary in time. At first, for simplicity, we study the effect of small perturbations. Then, mathematically, the central object of sensitivity analysis is the \emph{sensitivity matrix}. This matrix encodes the partial derivatives of the responsive components with respect to the parameters, at equilibrium. For a chemical network case, natural responsive components to be considered are the concentrations of chemicals, and the reaction fluxes.\\

In applications, precise measurements are often very difficult - if not impossible - and reaction rates remain largely unknown in most specific cases. For this reason, we aim for a comprehensive \emph{qualitative analysis} rather than quantitative numerical simulations for one or another set of guesswork parameters. Our qualitative analysis is based on the structure of the network, only.\\

Various types of sensitivity analysis are common in the frame of chemistry. We refer to the survey paper \cite{SRTC05} for more detailed references. An interesting approach, in a deterministic context, has been developed by Shinar, Feinberg, and co-authors \cite{ShiFei09, ShiFei10, ShiFei11}. In this body of work, the concept of \emph{absolute concentration robustness} (ACR) has been introduced. In the authors' words \cite{ShiFei11}, ``a model biochemical system has ACR relative to a particular bio-active molecular species if [...] the concentration of that species is the same in all of the positive steady states that the system might admit, regardless of the overall supplies of the various network constituents''. ACR thus indicates zero sensitivity of the concentration of a certain species with respect to the other network components. Moreover, in \cite{Shietal11} Shinar and co-authors were able to derive quantitative bounds on the entries of the sensitivity matrix for reaction fluxes, in a mass-action kinetics context and for a regular class of networks. Recently, and in an algebraic context,  Feliu also addressed related problems \cite{Feliu19}, focusing on polynomial systems.\\

In parallel, Fiedler and Mochizuki pursued a sensitivity analysis in a more metabolically oriented context \cite{MF15, FM15}, with applications to the central glucose metabolism of Escherichia Coli. These works, and especially the generalization with Brehm \cite{BF18}, are the starting point for the present work, which is a continuation of their contribution. In fact, the previous body of work was restricted to perturbations of reaction rates, and did not consider the case of metabolite concentrations perturbation, which is addressed for the first time in the present paper. Their main interest was to develop a mathematical theory able to model enzyme knock-out experiments on the central glucose metabolism of Escherichia Coli. The experimental paper for their reference was the fundamental contribution \cite{Ish07} by Ishii and co-authors, which investigated the responses of Escherichia Coli to genetic perturbations of knock-out type. The biological paper \cite{MF15} outlined the modeling approach and symbolically computed the responses for a model of the central metabolism of Escherichia Coli.\\

In practice, the computation of the sensitivity responses for a large metabolic network requires intense computational effort. However, interestingly, the responses computed at first in \cite{MF15} showed an unexpected and intriguing pattern feature, with a high number of zero responses (\emph{sparsity of sensitivity}) and interrelated responses.\\ 
The mathematical companion paper \cite{FM15} started from this intuition, highlighting the algebraic structures responsible for those patterns. The analysis, there, was restricted to the simpler case of monomolecular networks, that is, networks where each reaction transforms one single metabolite $m_1$ to one single metabolite $m_2$, only. This, in particular, allowed a full description in terms of directed graphs, see also Section \ref{monosec} of the present paper. The nonzero response of a metabolite $m'$ and a reaction $j'$, to a perturbation of the rate of a reaction $j^*$ has been called \emph{nonzero influence of $j^*$ on $m'$, $j'$}, and it has been denoted with the graphical representation
\begin{equation}
j^* \rightsquigarrow m' \quad, \quad j^* \rightsquigarrow j' \,.
\end{equation}
In particular, transitivity of flux influence 
\begin{equation}\label{introtrans}
j^* \rightsquigarrow {j'} \quad \text{and} \quad j' \rightsquigarrow j'' \quad \Rightarrow  \quad j^* \rightsquigarrow j'' 
\end{equation}
was established, for the monomolecular case. The transitivity statement \eqref{introtrans} looks deceptively simple but turned out to be a fairly delicate topic. In fact, the perturbation spreads also along other components of the network, and this effect needs to be taken in account. In particular, for example, the present paper shows, in Section \ref{trans}, how transitivity does \emph{not} hold in the case of metabolite influence for general multimolecular networks:
\begin{equation}
m^* \rightsquigarrow m' \quad \text{and} \quad m' \rightsquigarrow m''  \quad \not\Rightarrow  \quad m^* \rightsquigarrow m''.
\end{equation}

The pattern formation has been further studied by Okada and co-authors in \cite{OM16, OM17, OTM18}. Connections with the existing sensitivity and robustness theory, as developed by Shinar and co-authors, has been investigated by Sasha Siegmund in \cite{Sie16}. In a joint paper with Matano \cite{VM17}, we have given a more refined formulation of the monomolecular transitivity result of \cite{FM15}, essentially based on the classic Menger's theorem \cite{Men27}, and we have described the structure of the \emph{influenced sets} $\mathbf{I}(j^*):= \{j' : j^* \rightsquigarrow j' \}$. A first analysis for the signed response to reaction perturbations, in the monomolecular case, was established in \cite{V17}.\\

In 2018, Brehm and Fiedler \cite{BF18} addressed, with similar settings, the multimolecular case, again only considering reaction rates perturbations. They achieved an algebraic description of the responses and established transitivity of flux influence \eqref{introtrans} also for this general case. The main tool of the analysis in \cite{BF18} are the \emph{Child Selections}. A Child Selection map $\mathbf{J}$ is an injective map from the metabolite set $\mathbf{M}$ to the reaction set $\mathbf{E}$, associating to any input \emph{mother} metabolite $m$ an output \emph{child} reaction $j$, outgoing from $m$. A Child Selection identifies reshuffled square submatrices $S^{\mathbf{J}}$ of the stoichiometric matrix, whose $m^{th}$ column corresponds to the stoichiometric column of the reaction $j=\mathbf{J}(m)$. In addition, \cite{VGB20} concentrated and expanded on the language of Child Selections to investigate zero-eigenvalue bifurcations of equilibria. Child Selections, and in particular the submatrices $S^{\mathbf{J}}$, play a central role in the analysis of the present work, as well.\\

In this paper, the analysis of nonzero sensitivity has been completed to include the missing case of metabolite perturbation. Section \ref{metnet} and \ref{sensitivity} recall the mathematical settings of metabolic networks and sensitivity analysis, respectively. Section \ref{CSPCS} introduces Child Selections and Partial Child Selections, concepts needed for the main results. In a similar algebraic fashion to the language developed in \cite{BF18, VGB20}, Theorem \ref{metmet} describes the metabolite response to a metabolite perturbation. A particularly intriguing consequence of the theorem, Remark \ref{noself}, is that a metabolite $m$ does not always respond to a direct perturbation of the metabolite $m$ itself. We provide a network example of this fact, constructed by the intuition that Theorem \ref{metmet} provides. Respectively, Theorem \ref{fluxmet} focuses on the reaction flux response to a metabolite perturbation. Both theorems characterize the nonzero response with the existence of certain nonzero minors of the stoichiometric matrix, constructed with the language of Child Selections and Partial Child Selections.\\
As an important corollary to Theorems \ref{metmet}, \ref{fluxmet} and previous results \cite{BF18}, Section \ref{red} concludes that a metabolite perturbation can be reduced to a reaction perturbation, from a mathematical perspective. In fact: a perturbation of a metabolite $m^*$ corresponds identically to a perturbation of an inflow reaction to $m^*$. After having established this point, much of the mathematics related to the two types of perturbation can be developed jointly.\\
Section \ref{monosec} illustrates our analysis for the simpler class of monomolecular networks. The description is in terms of directed paths in the network. Theorem \ref{monomet} states that the response of an element $p$, either metabolite or reaction, to a metabolite perturbation of  $m^*$, is nonzero if and only if $p$ is reachable from $m^*$ via a directed path, in the usual graph theory sense.\\
Finally, Section \ref{trans} concentrates on the transitivity problem for the missing case of metabolite perturbation and shows, with a simple counterexample, that the Brehm-Fiedler result \cite{BF18} does not extend to metabolite perturbation. Section \ref{discussion} presents a final summarizing discussion and all proofs are listed in Section \ref{pfs}.\\

\smallskip 

\textbf{Acknowledgments:} \textit{We are grateful to Bernold Fiedler for his encouragement, advice and support. Thanks also to Bernhard Brehm for early discussions. This work has been supported by the Collaborative Research Center (SFB) 910 and the Berlin Mathematical School.}\\

\section{Metabolic networks in mathematics} \label{metnet}

A metabolic network $\mathbf{\Gamma}$ is a pair $\{ \mathbf{M},\mathbf{E}\}$, where \textbf{M} is a set of $|\mathbf{M}|=M$ metabolites and $\textbf{E}$ is a set of $|\mathbf{E}|=N$ reactions. In examples, we use labels $A, B, C, D, ...$ for metabolites and $1, 2, 3....$ for reactions. The small letter $m \in \mathbf{M}$ is used for a generic metabolite and the small letter $j \in \mathbf{E}$ for a generic reaction.\\

A reaction $j$ is a transformation:
\begin{equation} \label{reactionj}
 j: \quad s^{j}_1m_1+...+s^{j}_Mm_M \underset{j}{\longrightarrow} \tilde{s}^{j}_1m_1+...+\tilde{s}^{j}_Mm_M.
\end{equation}
with nonnegative stoichiometric coefficients $s^{j},\tilde{s}^j$. In a metabolic context, these coefficients are integer, and mostly 0 or 1. However, the theory developed here applies indistinctly for real $s^{j},\tilde{s}^j \in \mathbb{R}$.\\
The metabolites $m$ appearing at the left hand side of \eqref{reactionj} with nonzero stoichiometric coefficient $s^{j}_m \neq 0$ are called \emph{inputs} or \emph{reactants} of the reaction $j$. Similarly, on the right hand side, the metabolites $m$ with $\tilde{s}^{j}_m\neq 0$ are called \emph{outputs} or \emph{products} of $j$.\\
 %We say, conversely, that a reaction $j$ is \emph{outgoing} from the metabolite $m$ if $m$ is an input of reaction $j$. We say that a reaction $j$ is an \emph{ingoing} reaction of the metabolite $m$ if $m$ is an output of the reaction $j$.\\
Metabolic systems are designed to transform nutrients into energy, so that they are naturally open systems, exchanging chemicals with the outside environment by inflows and outflows. Here, \emph{inflow reactions} are reactions with no inputs ($s^j=0$) and \emph{outflow reactions} are reactions with no outputs ($\tilde{s}^j=0$).\\

There are many ways in which the above combinatorial structure can be graphically represented. We consider the metabolites as vertices and the reactions as directed arrows of the network, which is one natural representation widely used in chemistry, biology, and mathematics. For example, the reaction
\begin{equation}\label{firstpiceq}
j: \quad  A+2B \underset{j}{\longrightarrow} C,
\end{equation}
is represented as follows,
\begin{equation}\label{firstpic}
\end{equation}
\begin{center} 
\vspace{-1cm}
\includegraphics[scale=0.35]{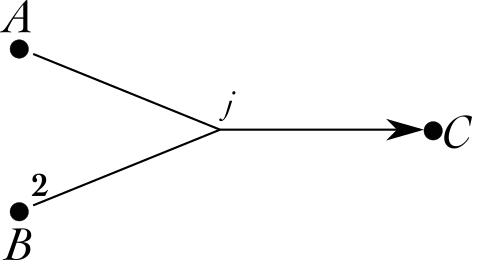}
\end{center}
with the stoichiometric coefficients different from $1$ appearing explicitly as a weight in the lower tail of the directed arrow $j$.\\

The stoichiometric matrix $S$ is the $M\times N$ matrix with entries
\begin{equation}\label{smatrix}
S_{mj}:=
\begin{cases}
-s^j_m\,\;\quad \text{ for $m$ input of $j$},\\
\tilde{s}^j_{m}\quad\quad \text{ for $m$ output of $j$},\\
0\quad\quad\quad\text{if $m$ does not participate in reaction $j$}.\\
\end{cases}
\end{equation}
For instance, in \eqref{firstpiceq}, let us assume the reaction $j$ appears in a network with four metabolites $\{A, B, C, D\}$. Such a reaction translates into the $j^{th}$ column of the stoichiometric matrix $S$ as
\begin{equation}
S^j=
\begin{blockarray}{cc}
 & j \\
\begin{block}{c(c)}
  A & -1\\
  B & -2\\
  C & 1\\
  D & 0\\
\end{block}
\end{blockarray}\;.
\end{equation}
Throughout, we always denote $S^j$ the column of the stoichiometric matrix $S$ associated to the reaction $j$. Note also that, in particular, we model a reversible reaction
\begin{equation}
j: \quad A \underset{j}{\longleftrightarrow} B+C
\end{equation}
simply as two irreversible reactions
\begin{equation}
j_1: \quad A \underset{j_1}{\longrightarrow} B+C\quad \text{ and }\quad j_2: \quad B+C \underset{j_2}{\longrightarrow} A.
\end{equation}
It follows from our construction that stoichiometric columns associated to inflow reactions have only positive entries and columns associated to outflow reactions have only negative entries.\\

Under the assumption that the reactor is isothermal and spatially homogeneous, the dynamics of the system is governed by the following system of $M$ ordinary differential equations (ODEs):
\begin{equation} \label{eq}
\dot{x}=f(x):=S \mathbf{r}(x),
\end{equation}
where $x_m(t)$ describes the time evolution of the concentration of the metabolite $m$. The vector $x(t)$ is considered positive, $x(t)>0$, since no physical meaning is given to a negative concentration. The $M \times N$ matrix $S$ is the stoichiometric matrix \eqref{smatrix}, and the $N$-dimensional vector $\mathbf{r}(x)$ represents the reaction rates (\emph{kinetics}) as functions of the concentrations.\\

Staying in wide generalities, we do not prescribe a precise form to the kinetics $\mathbf{r}(x)$. Instead, we simply consider any $r_j(x)$ as a positive $C^1$ function which depends only on the concentrations $x_m$ of those metabolites $m$ that are input to reaction $j$. For the special case of inflow `feed' reactions $j_f$, with no inputs, we consider the reaction rate as constant, that is
\begin{equation}
r_{j_f}(x)\equiv K_{j_f}.
\end{equation}
In the case in which $m$ is an input metabolite to reaction $j$, we use the notation $r_{jm}$ for the nonzero partial derivatives
\begin{equation}
r_{jm}:= \frac{\partial r_j(x)}{\partial x_m} \neq 0.
\end{equation}

Notorious examples that fall within the above class are \emph{mass action} kinetics and \emph{Michaelis-Menten} kinetics, among many others. Mass action kinetics (1869) considers the rate $r_j$ as polynomials in the concentration:
\begin{equation}\label{massaction}
r_j(x)=a_j \prod_{m\in\mathbf{M}} x_m^{s^j_m},
\end{equation}
where $a_j$ is a positive coefficient. Michaelis-Menten kinetics (1913), designed to model enzymatic reactions and more suited in a metabolic context, has the following rational form:
\begin{equation} \label{mmenten}
r_j(x)=b_j\prod_{m\in\mathbf{M}} \Bigg( \frac{x_m}{(1+c_j^m x_m)}\Bigg)^{s^j_m},
\end{equation}
where again all coefficients $b_j, c_j^m$ are considered positive. Note that, for $c_j^m \rightarrow 0$, we recover the mass action case \eqref{massaction}. Chemical kinetics is a vast field of study, which does not at all conclude here. See the encyclopedia \cite{CCK} for much more extended information.\\
Often, mass action kinetics is considered the simplest case, mathematically, due to its polynomial form. However, in our approach, we shall disagree with such a consideration. In fact, certain conclusions of the theory here presented result weaker in the mass action case. We expand on this in Section \ref{sensitivity} and in particular in the discussion about \emph{algebraically nonzero} statements.\\

The question about existence and uniqueness of an equilibrium solution $\bar{x}$ of \eqref{eq}, given the network structure and a kinetics class, have been addressed in wide strains of the literature, see for example fundamental concepts such as \emph{deficiency} \cite{Fei87}, \emph{injectivity} \cite{GaNi65, BaCra10}, and \emph{concordance} \cite{ShiFei12}. Interested in the sensitivity of \emph{existing} equilibria, we jump over these relevant questions, here. Instead, we \emph{assume} the existence of a positive equilibrium $\bar{x}$ solving
\begin{equation} \label{maineqeq}
0=S\mathbf{r}(\bar{x}).
\end{equation}

In the following Section \ref{sensitivity}, we address perturbations of the equilibrium system \eqref{maineqeq}, and the relative responses of the network.

\section{Sensitivity setting} \label{sensitivity}

For metabolic networks, sensitivity studies the response to external perturbations of the equilibrium system \eqref{maineqeq}. We consider the following perturbation of \eqref{maineqeq}:
\begin{equation} \label{perteq}
0=S\mathbf{r}(\bar{x})+\varepsilon e_{m^*} ,
\end{equation}
where $e_{m^*}$ is the $m^{*th}$ unit vector in $\mathbb{R}^M$. The perturbed equation \eqref{perteq} models a targeted perturbation of the single metabolite $m^*$. For simplicity of presentation, in this paper we focus on the case of targeted perturbations of single metabolites. In fact, any general perturbation vector $\mu \in \mathbb{R}^M$ can be clearly written as a linear combinations of targeted perturbations of the above form \eqref{perteq}. Similarly, the response to a general $\mu$-perturbation is again the linear combination of the responses to targeted perturbations. For further details see the doctoral thesis \cite{Vas20}, where the same theory is presented in more general settings.\\

%We address the sensitivity question at its basics:
%$$\textit{Which components of the network respond - qualitatively - to external perturbations?}$$
%That is:
%\begin{equation} \label{sensq}
%\textit{Which components of the network do respond, at all?}
%\end{equation}

Our object of study are the responses of the network to $\varepsilon$-small perturbations at a positive equilibrium $\bar{x}$. Vaguely, this means that we look at the algebraic form of the differentiated components of equation \eqref{perteq}, with respect to $\varepsilon$, at $\varepsilon=0$. There are only two `responsive' components: the metabolite concentrations $x$ themselves and the reaction fluxes $\mathbf{r}(x)$. Therefore, we consider:\\
\begin{enumerate}[itemsep=0pt]
\item the response vector of the metabolite concentrations to a perturbation of $m^*$:
\begin{equation} \label{concresp}
\delta x ^{m^*} := \frac{\partial \bar{x}}{\partial \varepsilon} \bigg|_{\varepsilon=0};
\end{equation}
\item the response vector of the reaction fluxes to a perturbation of $m^*$:
\begin{equation} \label{fluxresp}
\Phi^{m^*} := \frac{\partial \mathbf{r}(\bar{x})}{\partial \varepsilon} \bigg|_{\varepsilon=0} = R \; \delta x^{m^*}.
\end{equation}
\end{enumerate}
Here, $R$ is the $N \times M$ matrix of the partial derivatives $r_{jm}$:\\
\begin{equation} \label{Req}
R_{jm}:=\frac{\partial}{\partial x_m} r_j(x) =
\begin{cases}
r_{jm}\quad \text{if}\quad\frac{\partial r_j(x)}{\partial x_m}\neq 0\\
0\quad\quad\text{otherwise}
\end{cases}.
\end{equation}

%REWRITE AFTER SETTINGS Here, \textit{algebraically} means precisely as a rational function of the partial derivatives $r_{jm}$. It is crucial to clarify this as much as possible, with no fear of being pedantic. In fact, the responses are rational functions of the partial derivatives $r_{jm}$, as the continuation of this chapter shows. Non-identically zero rational functions of parameters may be zero for some values of the parameters. \textit{Algebraically nonzero} means that the rational function itself of those parameters is non-identically zero. Since our analysis is qualitative, we will not be able to predict a quantitative zero for certain specific reaction rates. Rather, we are only interested in the zeros for all parameters. All statements about any  response $\delta x^{m^*}$ and $\Phi^{m^*}$ must be intended in this algebraic sense, even if we omit to specify it.\\

The core tool of analysis is the Implicit Function Theorem (IFT), whose application requires a nondegeneracy assumption to be satisfied: the Jacobian matrix of the unperturbed equilibrium system \eqref{maineqeq}, $\frac{\partial}{\partial x}S\mathbf{r}(x)=SR$, needs to be invertible. That is, 
\begin{equation} \label{nd}
\operatorname{det}SR\neq 0.
\end{equation} 
We assume \eqref{nd}, throughout, and we call \emph{nondegenerate} a metabolic network $\mathbf{\Gamma}$ for which \eqref{nd} is satisfied. For metabolic networks, this condition is not particularly restrictive. Section \ref{CSPCS}, Corollary \ref{cornondeg}, provides a structural characterization of a network being nondegenerate.\\
%This assumption excludes conserved quantities, that is, left kernels of the stoichiometric matrix. Throughout the paper, we assume the condition \ref{nd}. It is not a restrictive assumption in our case.  A network characterization of such nondegeneracy has been described in \cite{BF18,VGB20}. 

Under \eqref{nd}, the IFT guarantees the existence of a family of equilibrium solutions $\bar{x}(\varepsilon)$ to equation \eqref{perteq}, for sufficiently small $\varepsilon$. In particular, the following equality holds, by differentiation:
\begin{equation} \label{diffeq}
0=\frac{\partial}{\partial \varepsilon} (S \mathbf{r}(\bar{x}(\varepsilon)) + \varepsilon e_{m^*}) = S R \; \delta x^{m^*}+e_{m^*},
\end{equation}
which leads to the equalities:
\begin{equation} \label{mmet}
\delta x^{m^*} = -(SR)^{-1} e_{m^*},
\end{equation}
and, via \eqref{fluxresp},
\begin{equation}\label{fflux}
\Phi^{m^*}=-R(SR)^{-1} e_{m^*}.
\end{equation}

%Above, the matrix of the responses $\delta x$ of metabolite concentrations to a perturbation of concentrations are identified with the inverse of the Jacobian matrix, with opposite sign. This is in accordance with the \emph{ecology community},  which studied similar problems. See for example \cite{Yo88} and \cite{Na92}, where the sensitivity matrix for `\emph{food webs}' and `\emph{flow networks}' has been studied.\\

In our approach, we interpret \eqref{mmet} and \eqref{fflux} as \emph{functions of the partial derivatives $r_{jm}$}. Consequently, the response $\delta x^{m^*}_{m'}$ of metabolite $m'$ (respectively $\Phi^{m^*}_{j'}$ of reaction $j'$) is termed \emph{algebraically nonzero} if it is non identically zero as a function of the derivatives $r_{jm}$. Conversely, an identically zero function is called \emph{zero response}. Moreover, we say that a metabolite $m'$ (or a reaction $j'$) is \emph{influenced} by ${m^*}$ if $(\delta x)^{m^*}_{m'} \neq 0$ ($(\Phi)^{m^*}_{j'}\neq 0$, resp.), algebraically, and we denote this by
\begin{equation}
{m^*}\rightsquigarrow m' \quad ({m^*}\rightsquigarrow j',\;\text{resp.}).
\end{equation}

Thus, crucially, we consider the derivatives $r_{jm}$ as \emph{parameters} for the analysis. An algebraically zero response strictly implies that the response is zero for any parameter in any class of kinetics. On the other hand, of course, an algebraically nonzero response does not exclude that the response may actually be evaluated zero for \emph{certain} values of $r_{jm}$. The underlying question whether, given a parametric kinetics class, we may always and automatically consider these parameters $r_{jm}$ independent from each other and from the equilibrium constraints \eqref{maineqeq} is significant. A positive answer guarantees indeed, without further analysis, the existence of certain parameters at which an algebraically nonzero response can also be evaluated nonzero. The answer strongly depends on the class of kinetics we are considering. For example, for mass action kinetics, at a fixed equilibrium $\bar{x}$, the value $r_j(\bar{x})=k_j \bar{x}$ determines also the value of the derivative $r_{jm}=k_j=\frac{r_j(\bar{x})}{\bar{x}}$. In particular, we cannot consider the derivatives $r_{jm}$ independent from the equilibrium constraints \eqref{maineqeq}. Already the slightly reacher class of Michaelis-Menten kinetics provide enough parametric freedom to consider the derivatives $r_{jm}$ as free parameters, independent from each other and from the equilibrium \eqref{maineqeq}. The explicit derivation has been presented in \cite{VGB20} and we omit it here. In this latter Michaelis-Menten case, we can directly conclude that an algebraically nonzero response implies that there exist reaction rates parameters at which the response can be evaluated nonzero. On the contrary, in the former mass action case, further analysis is needed. In this sense, in our approach, a parametrically `poor' kinetics as mass action is more limiting than parametrically richer kinetics.\\

Already for reasonably small networks, the complexity of the explicit computation of $(SR)^{-1}$, symbolically, in terms of the derivatives $r_{jm}$ is far too high and unfeasible as a routine method of analysis. Hence, one of the main goals of our approach in general, and this paper in particular, is to simplify the computation, by algebraic characterizations that circumvent the brutal symbolic inversion of the Jacobian matrix $SR$. Theorems \ref{metmet} and \ref{fluxmet} structurally characterize expressions \eqref{mmet} and \eqref{fflux} to be algebraically nonzero according to certain minors of the stoichiometric matrix $S$. Since the stoichiometric entries are integers, and non symbolic, this leads to computational simplification as well as meaningful insights.\\

The following Section \ref{CSPCS} recalls the language of Child Selections and Partial Child Selections.

\section{Child Selections and Partial Child Selections} \label{CSPCS}

We start this section by recalling some definitions from \cite{BF18,VGB20}.

\begin{defn}[Child Selections, mothers, children]
A \emph{Child Selection} is an injective map $\mathbf{J}: \textbf{M} \longrightarrow \textbf{E}$, which associates to every metabolite $m \in \textbf{M}$ a reaction $j \in \textbf{E}$ such that $m$ is an input metabolite of reaction $j$. We call the reaction $j=\mathbf{J}(m)$ \emph{child} of $m$, and the metabolite $m=\mathbf{J}^{-1}(j)$ \emph{mother} of the reaction $j$.
\end{defn}

As analyzed in \cite{BF18,VGB20}, the Jacobian determinant of $SR$ can be expressed in terms of Child Selections $\mathbf{J}$:
\begin{equation} \label{vita}
\operatorname{det}SR=\sum_\mathbf{J} \operatorname{det}S^\mathbf{J} \cdot \prod_{m\in \mathbf{M}} r_{\mathbf{J}(m)m},
\end{equation}
where $S^\mathbf{J}$ is the matrix whose $m^{th}$ column is the $\mathbf{J}(m)^{th}$ column of $S$. In particular, the columns of $S^\mathbf{J}$ correspond one-to-one, and following the order, to the reactions
$$\mathbf{J}(m_1), \; \mathbf{J}(m_2), \;...\; , \;\mathbf{J}(m_{M-1}), \;\mathbf{J}(m_{M}).$$

A network characterization of the required nondegeneracy assumption \eqref{nd}, in our algebraic sense, follows as a straightforward corollary to \eqref{vita}.
\begin{cor} \label{cornondeg}
A metabolic network is \emph{nondegenerate}, i.e.
$$\operatorname{det}SR \neq 0,\text{ algebraically},$$
if and only if there exists a Child Selection $\mathbf{J}$, such that 
\begin{equation}
\operatorname{det}S^\mathbf{J} \neq 0.
\end{equation}
\end{cor}
We can already see here a major computational simplification. In fact, the question whether the Jacobian determinant of $SR$ is algebraically nonzero is of high complexity due to the symbolic entries of $SR$. However, we are able to answer this question just with the existence of a nonsingular submatrix $S^\mathbf{J}$ of the stoichiometric matrix $S$. Good circumstance: the stoichiometric matrix $S$ is integer valued and consequently any computation involving $S$ is much faster.\\

To state our results we introduce a further related concept: the \textit{Partial Child Selections} (PCS).

\begin{defn} [Partial Child Selections]
A Partial Child Selection $\mathbf{J^{\vee m_i}}$ is an injective map from the metabolite set $\mathbf{M} \setminus \{m_i\}$ to the reaction set $\mathbf{E}$, such that to any metabolite $m \neq m_i$ is associated an outgoing reaction of $m$, injectively.
\end{defn}

Without loss of generality, assume $1, ... , i, ... , M$. In analogy to the submatrix $S^{\mathbf{J}}$ for a Child Selection $\mathbf{J}$, the expression $S^{\mathbf{J^{\vee m_i}}}$ indicates then, for a Partial Child Selection $\mathbf{J^{\vee m_i}}$, the $M \times (M-1)$ matrix with columns corresponding one-to-one, and following the order, to the reactions
$$\mathbf{J^{\vee m_i}}(m_1),\;...\;,\;\mathbf{J^{\vee m_i}}(m_{i-1}), \;\mathbf{J^{\vee m_i}}(m_{i+1}),\;...\;,\; \mathbf{J^{\vee m_i}}(m_M).$$
That is, the first column is the stoichiometric column $S^{j_1}$ of the reaction $j_1=\mathbf{J^{\vee m_i}}(m_1)$. Analogously, the $i^{th}$ column is the stoichiometric column $S^{j_i}$ of the reaction $j_i=\mathbf{J^{\vee m_i}}(m_{i+1})$, and so on. 

%Finally, the notation $S^{\mathbf{J^{\vee m_i}}}_{\vee m^*}$ indicates the $(M-1)\times(M-1)$ square matrix obtained from $S^{\mathbf{J^{\vee m_i}}}$ by removing the $m^{*\;th}$ row.\\

%In general, the interpretation of the square matrices $S^{\mathbf{J^{\vee m_i}}}_{\vee m^*}$ in terms of subnetworks is rather abstract and it is not advantageous, so we do not linger on it, here. It is good to mention, though, the special case of $m_i = m^*$. In this case, the algebraic structure of $S^{\mathbf{J^{\vee m^*}}}_{\vee m^*}$ is identical to the algebraic structure of $S^\mathbf{J}$.
\begin{rmk} \label{framermk} We point at a deceptive feature of Partial Child Selections. A Partial Child Selection may innocently look as a restriction of Child Selections, in the sense that from each Partial Child Selection $\mathbf{J^{\vee m_i}}$ it may be possible to induce an associated Child Selection $\mathbf{J}$ such that $\mathbf{J}(m)=\mathbf{J^{\vee m_i}}(m)$ for any $m \neq m'$. This is actually \emph{not} always the case. For example, let us consider the network:
\begin{equation} \label{partialchild}
\end{equation}

\vspace{-0.7cm}

\begin{center} 
\includegraphics[scale=0.3]{PARTIALCHILD.png}
\end{center}

with stoichiometric matrix
\[
S=
\begin{blockarray}{cccc}
 & 1 & 2 & 3\\
\begin{block}{c[ccc]}
  A & -1 & 0 & 0\\
  B & -1 & -1 & 0\\
  C & 1 & 0 & -1\\
\end{block}
\end{blockarray}\;.
\]

We have omitted here inflow reactions, as they are superfluous in the argument. In \eqref{partialchild} there is only one Child Selection, namely $\mathbf{J}:=\{ \mathbf{J}(A)=1; \mathbf{J}(B)=2; \mathbf{J}(C)=3\}$. However, we have \emph{two} Partial Child Selections $\mathbf{J_1^{\vee A}}$ and $\mathbf{J_2^{\vee A}}$ on the set $\{ B, C \}$, that is, with vertex $A$ being removed. They are:
\begin{enumerate}[itemsep=0pt]
\item $\mathbf{J_1^{\vee A}}:=\{ \mathbf{J_1^{\vee A}}(B)=1; \mathbf{J_1^{\vee A}} (C)=3\}$;
\item $\mathbf{J_2^{\vee A}}:=\{\mathbf{J_2^{\vee A}}(B)=2; \mathbf{J_2^{\vee A}} (C)=3\}$.
\end{enumerate}
Via a Child Selection $\mathbf{J}$, $B$ must select reaction $2$, due to injectivity, because the reaction $1$ must be chosen by $A$. In contrast, via a Partial Child Selection $\mathbf{J^{\vee A}}$, $B$ can freely select both reactions $1$ and $2$.\\ 
\end{rmk}

\section{Metabolite concentrations response} \label{secmetmet}

We analyze the expression
\begin{equation}
(\delta x)^{m^*}_{m'}=-((SR)^{-1})^{m^*}_{m'},
\end{equation}
which describes the response of the concentration of $m'$ to a perturbation of the concentration of $m^*$. Here, and in the next sections, $((SR)^{-1})^{m^*}_{m'}$ indicates the entry of the matrix $(SR)^{-1}$ in the $m^{*\;th}$ column and $m'^{\;th}$ row. Let the expression
\begin{equation} \label{not1}
S^{\mathbf{J^{\vee m'}\cup e_{m^*}}}
\end{equation}
indicate the $M\times M$ matrix obtained from $S^{\mathbf{J^{\vee m'}}}$ by inserting, as new $m'\;^{th}$ column, the unit vector $e_{m^*}\in \mathbb{R}^M$. In particular, the columns of $S^{\mathbf{J^{\vee m'}\cup e_{m^*}}}$ correspond one-to-one, and following the order, to
$$\mathbf{J^{\vee m_i}}(m_1),\;...\;,\;\mathbf{J^{\vee m_i}}(m_{i-1}), \quad e_{m^*},\quad \mathbf{J^{\vee m_i}}(m_{i+1}),\;...\;,\; \mathbf{J^{\vee m_i}}(m_M).$$
The result for the metabolite response to a metabolite perturbation reads as follows:

\begin{thm}\label{metmet}
Let $m^*$ and $m'$ be two (not necessarily distinct) metabolites. Then the response $(\delta x)^{m^*}_{m'}$ of metabolite $m'$ to a targeted perturbation of the metabolite $m^*$ is given by the formula:
\begin{equation} \label{Bformulametmet}
(\delta x)^{m^*}_{m'}=-\frac{\sum \limits_{\mathbf{J^{\vee m'}}} \operatorname{det} S^{\mathbf{J^{\vee m'}\cup e_{m^*}}} \cdot \prod_{m\in \mathbf{M}\smallsetminus {m'}} r_{{\mathbf{J^{\vee m'}}(m)m}}}{\operatorname{det}SR}.
\end{equation}
In particular,
 $$(\delta x)^{m^*}_{m'}\neq 0,\text{ algebraically,}$$
if, and only if, there exists a Partial Child Selection 
$$\mathbf{J^{\vee m'}}:  \mathbf{M}\smallsetminus \{m' \} \longrightarrow \mathbf{E}$$
such that 
\begin{equation}
 \operatorname{det} S^{\mathbf{J^{\vee m'}\cup e_{m^*}}} \neq 0.
\end{equation}
\end{thm}

Note that the expression \eqref{Bformulametmet} is a rational function in the variables $r_{jm}$, whose numerator is a homogenous multilinear polynomial of degree $M-1$, and whose denominator is a homogenous multilinear polynomial of degree $M$ (the Jacobian determinant of $SR$).

\begin{rmk}\label{noself}
One important and counterintuitive consequence of Theorem \ref{metmet} is that self-influence $m^* \rightsquigarrow m^*$ may not happen. Theorem \ref{metmet} provides also an intuition about how this peculiar case occurs: it is the case when all Partial Child Selections $\mathbf{J^{\vee{m^*}}}$ identify singular matrices $S^{\mathbf{J^{\vee m^*}\cup e_{m^*}}}$. The following network example is one of such cases.
\begin{equation}
\end{equation}

\vspace{-0.5cm}

\begin{center} 
\includegraphics[scale=0.38]{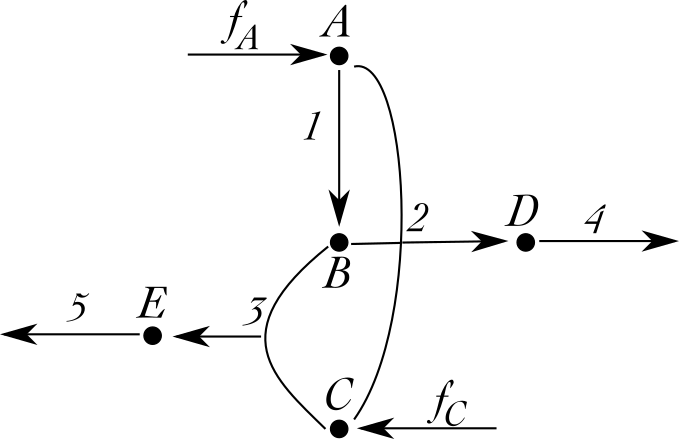}
\end{center}
with stoichiometric matrix S and reaction rates $\mathbf{r}(x)$:
\[
S=
\begin{blockarray}{cccccccc}
 & f_A & f_C & 1 & 2 & 3 & 4 & 5\\
\begin{block}{c[ccccccc]}
  A & 1 & 0 & -1 & -1 & 0 & 0 & 0\\
  B & 0 &  0 & 1&  -1 & -1 & 0 & 0\\
  C & 0 & 1 &  0 & -1 & -1 & 0 & 0\\
  D & 0 & 0 &  0 & 1 & 0 & -1 & 0\\
  E & 0 & 0 &  0 & 0 &  1 & 0 & -1\\
\end{block}
\end{blockarray}; \quad \quad \quad \quad \quad
\mathbf{r}(x)=
\begin{blockarray}{c}
\begin{block}{(c)}
K_{f_A}\\
K_{f_C}\\
r_1(x_A)\\
r_2(x_A, x_B, x_C)\\
r_3(x_B, x_C)\\
r_4(x_D)\\
r_5(x_E)\\
\end{block}
\end{blockarray}.
\]

Note that reactions $f_A$ and $f_C$ are inflows to metabolites $A$ and $C$, respectively. Reactions $r_4$ and $r_5$ are outflows from metabolites $D$ and $E$, respectively. The dynamical system has the following explicit form
\[
\begin{blockarray}{c}
\begin{block}{(c)}
\dot{x}_A\\
\dot{x}_B\\
\dot{x}_C\\
\dot{x}_D\\
\dot{x}_E\\
\end{block}
\end{blockarray}=
S\mathbf{r}(x)=
\begin{cases}
K_{f_A}-r_1(x_A)-r_2(x_A, x_B, x_C)\\
r_1(x_A)-r_2(x_A, x_B, x_C)-r_3(x_B, x_C)\\
K_{f_C}-r_2(x_A, x_B, x_C)-r_3(x_B, x_C)\\
r_2(x_A, x_B, x_C) - r_4(x_D)\\
r_3(x_B, x_C) - r_5 (x_E)\\
\end{cases}.
\]
The system admits an equilibrium for certain reaction rates. Moreover, there exists a Child Selection $\mathbf{J}$,
$$\mathbf{J}(A,B,C,D,E)\longrightarrow(1,2,3,4,5)$$
with 
\[
\operatorname{det}S^{\mathbf{J}}=
\operatorname{det}
\begin{blockarray}{ccccc}
\begin{block}{[ccccc]}
 -1 & -1 & 0 & 0 & 0\\
 1&  -1 & -1 & 0 & 0\\
 0 & -1 & -1 & 0 & 0\\
  0 & 1 & 0 & -1 & 0\\
  0 & 0 &  1 & 0 & -1\\
\end{block}
\end{blockarray}\; = -1 \neq 0.
\]
Thus, via Corollary \ref{cornondeg}, the system is nondegenerate and we may apply our sensitivity theory. The self-influence of metabolite $A$, namely the response of the concentration of metabolite $A$ to a perturbation of the concentration of $A$ itself, is algebraically nonzero if and only if there exists a Partial Child Selection 
$$\mathbf{J^{\vee A}}:  \mathbf{M}\smallsetminus \{A\} \longrightarrow \mathbf{E}$$
 such that 
 $$\operatorname{det} S^{\mathbf{J^{\vee A}\cup e_{A}}} \neq 0.$$
 
In the above example there are only two PCS $\mathbf{J^{\vee A}}$, namely
\begin{enumerate}
\item $\mathbf{J_1^{\vee A}}(B, C, D, E) \longrightarrow (2, 3, 4, 5)$;
\item $\mathbf{J_2^{\vee A}}(B, C, D, E) \longrightarrow (3, 2, 4, 5)$,
\end{enumerate}
with 
\[
S^{\mathbf{J_1^{\vee A} \cup e_{A}}}=
\begin{blockarray}{cccccc}
 & e_A & 2 & 3 & 4 & 5\\
\begin{block}{c[ccccc]}
A & 1 & -1& 0 & 0 & 0\\
B & 0 & -1 & -1 & 0& 0\\
C & 0 & -1 & -1 & 0 & 0\\
D & 0 & 1 & 0 & -1 & 0\\
E & 0 & 0 & 1 & 0 & -1\\
\end{block}
\end{blockarray},\quad \quad \text{and} \quad \quad 
S^{\mathbf{J_2^{\vee A} \cup e_{A}}}=
\begin{blockarray}{cccccc}
& e_A & 3 & 2 & 4 & 5\\
\begin{block}{c[ccccc]}
A & 1 & 0 & -1 & 0 & 0\\
B & 0 & -1 & -1 & 0& 0\\
C & 0 & -1 & -1 & 0 & 0\\
D & 0 & 0 & 1 & -1 & 0\\
E & 0 & 1 & 0 & 0 & -1\\
\end{block}
\end{blockarray},
\]

By Laplace expansion along the columns 1, 4, and 5, we clearly see that both matrices are singular. 
$$
\operatorname{det}S^{\mathbf{J_1^{\vee A} \cup e_{A}}}=\operatorname{det}S^{\mathbf{J_2^{\vee A} \cup e_{A}}}= \operatorname{det}\begin{blockarray}{cc}
\begin{block}{[cc]}
   -1 & -1 \\
  -1 & -1 \\
\end{block}
\end{blockarray}=0.
$$
We conclude that there is no self-influence of $A$:
$$A \not\rightsquigarrow A. $$

\end{rmk}

\section{Reaction fluxes response} \label{secfluxmet}

%Expression \eqref{tico} indicates that we have to compute an inner product between the row vector $R_{j'}$ and the column vector $-((SR)^{-1})^{m^*}$. We know both vectors: $R_{j'}$ by the network structure and $-((SR)^{-1})^{m^*}$ by Theorem \ref{metmet}. It is possible to prove a result about the flux response analyzing this inner product. However, we take here an independent route, so that any of our results can be considered singularly. 

We analyze the expression
\begin{equation} \label{tico}
(\Phi)^{m^*}_{j'}=(R\;\delta x)^{m^*}_{j'}=R_{j'}(\delta x)^{m^*}=-R_{j'}((SR)^{-1})^{m^*}.
\end{equation}
which describes the response of the flux of the reaction $j'$ to a perturbation of the concentration of $m^*$. Let now the expression
\begin{equation} \label{not2}
S^{\mathbf{J} \setminus  j' \cup {e_{m^*}}}
\end{equation}
indicate the $M \times M$ matrix obtained from $S^\mathbf{J}$ by removing the column $S^{j'}$, for $j'\in \mathbf{J}$, and replacing it with the unit vector $e_{m^*}\in \mathbb{R}^M$, in the same position of  $S^{j'}$.\\ 

Our result again characterizes the flux response to metabolite perturbations in terms of Child Selections.

\begin{thm}\label{fluxmet}
Let $m^*$ be a metabolite and $j'$ be a reaction. Then the flux response $(\Phi)^{m^*}_{j'}$ of reaction $j'$ to a targeted perturbation of the concentration of metabolite $m^*$ is given by
\begin{equation} \label{Bformulafluxmet}
(\Phi)^{m^*}_{j'}=-\frac{\sum \limits_{\mathbf{J}\ni j'} \operatorname{det}S^{\mathbf{J} \setminus  j' \cup {e_{m^*}}}  \prod_{m\in \mathbf{M}} r_{\mathbf{J}(m),m}}{\operatorname{det}SR}.
\end{equation}
In particular,
$$(\Phi)^{m^*}_{j'}\neq 0, \text{ algebraically,}$$
if, and only if, there exists a Child Selection $\mathbf{J}$ containing $j'$ and such that 
\begin{equation}
\operatorname{det}(S^{\mathbf{J} \setminus  j' \cup {e_m{^*}}})\neq 0.
\end{equation}
\end{thm}
The expression \eqref{Bformulafluxmet} is a rational function, whose both numerator and denominator are homogenous multilinear polynomials of degree $M$.

\begin{rmk}\label{inflowresp}
The response of an inflow reaction $j_f$ is always zero. This can be trivially seen by the fact that the reaction rate of $r_{j_f}$ is constant. Alternatively and consistently, note that there are no Child Selections $\mathbf{J}$ such that $j_f \in \mathbf{J}$, since $j_f$ has no input. Therefore Theorem \ref{fluxmet} concludes that $(\Phi)^{m^*}_{j_f}\equiv 0$ for any $m^*$.
\end{rmk}

\section{Reducing metabolite perturbations to reaction perturbations} \label{red}

In \cite{BF18}, Brehm and Fielder analyzed the response to \emph{reaction rates perturbations} in a parallel fashion to the present paper. A very brief recall of their work: they considered the following perturbed equilibrium equation:
\begin{equation} \label{reacperteq}
0=S\mathbf{r^\varepsilon}(\bar{x}),\quad\quad\quad \text{where} \quad \quad \quad \mathbf{r^\varepsilon}=\mathbf{r}+\varepsilon e_{j^*}.
\end{equation}
Above, $e_{j^*}$ is the unit vector in $\mathbb{R}^N$. Equation \eqref{reacperteq} models a targeted perturbation of the rate of the reaction $j^*$, exactly as our perturbed equation \eqref{perteq} models a targeted perturbation of the concentration of the metabolite $m^*$. Under the same nondegeneracy assumption \eqref{nd}, the concentration response $\delta x^{j^*}$ and the flux response $\Phi^{j^*}$ are defined analogously as in \eqref{concresp} and \eqref{fluxresp}, following the same approach of the present paper. In particular the following structural equalities, parallel to \eqref{mmet} and \eqref{fflux}, hold
\begin{enumerate}
\item $\delta x^{j^*}= -(SR)^{-1}S \; e_{j^*}$;
\item $\Phi^{j^*}=(Id_N-R(SR)^{-1}S)e_{j^*}$.
\end{enumerate}
Above, $Id_N$ is the $N$-dimensional identity matrix, $e_{j^*}$ is the unit vector in $\mathbb{R}^N$, $S$ and $R$ are again the stoichiometric matrix \eqref{smatrix} and the  matrix \eqref{Req} of the partial derivatives.\\

The main conceptual observation we make is the following. Consider an inflow reaction $f_{m^*}$ to a single metabolite $m^*$, i.e., $f_{m^*}$ possesses no inputs and the only product is $m^*$:
\begin{equation} \label{feedmstar}
f_{m^*}: \quad \quad \quad  \underset{f_{m^*}} {\longrightarrow}  \quad m^*.
\end{equation} 
Then, the stoichiometric column $S^{f_{m^*}}$ associated to $f_{m^*}$ \emph{corresponds} to the unit vector $e_{m^*} \in \mathbb{R}^M$ of the formulas \eqref{Bformulametmet} and \eqref{Bformulafluxmet}  appearing in Theorem \ref{metmet} and \ref{fluxmet}, respectively. Intuitively, this provides a connection between the perturbation of a metabolite concentration and of a reaction. Following this observation, we present a corollary to the theory presented in the present paper and in \cite{BF18}. The corollary, in particular, serves as a unification of the two approaches.

\begin{cor}\label{corred}
Let $f_{m^*}$ be an inflow reaction to metabolite $m^*$. Then, we have the following equalities:
\begin{equation}
(\delta x)^{m^*}_{m'}=(\delta x)^{f_{m^*}}_{m'}, \quad \quad \quad \text{for any $m'$};
\end{equation}
and 
\begin{equation}
\Phi^{m^*}_{j'}=\Phi^{f_{m^*}}_{j'}, \quad \text{for $j' \neq f_{m^*}$.}
\end{equation}
That is, the responses to a metabolite perturbation of the metabolite $m^*$ correspond to a reaction perturbation of an inflow reaction $f_{m^*}$ to $m^*$.
\end{cor}

\begin{rmk}
Note again that the case $j'=f_{m^*}$ is uninteresting, since $f_{m^*}$ is an inflow reaction and therefore $\Phi^{m^*}_{f_{m^*}}\equiv 0$ (Remark \ref{inflowresp}).
\end{rmk}
\begin{rmk}
Corollary \ref{corred} implies that, abstractly, reaction perturbations and metabolite perturbations are not two distinct mathematical questions. This mathematical fact clearly does not depend on the very fact that a metabolite $m^*$ possesses or not an inflow reaction. In further mathematical inquiries, such as the sign of the responses, there is thus no need to develop an independent theory for metabolite perturbations, as it is already included in the case of reaction perturbations.
\end{rmk}

\begin{rmk}
A careful reader might wonder whether it is possible to prove Corollary \ref{corred} independently from Theorems \ref{metmet} and \ref{fluxmet}. In this way, these two theorems would actually be just corollaries themselves of \ref{corred}. This is possible indeed. However, arguing in that way would hamper the self-containment of our presentation, and it would strongly require the reader to know the details of the reaction perturbation setting presented in \cite{BF18}, only sketched in this section. Preferring self-contained arguments, we have therefore decided to prove Theorems \ref{metmet} and \ref{fluxmet} independently, here, rather than making them follow from previous results.
\end{rmk}

\section{The monomolecular case} \label{monosec}

A monomolecular reaction network consists of metabolites $m$, which interact singularly by certain reactions $j$. That is, a monomolecular reaction network possesses only monomolecular reactions $j$ of the form
\begin{equation} \label{monomoleq}
j: \quad m_1 \underset{j}{\longrightarrow} m_2,
\end{equation}
where one single metabolite input $m_1$ is converted into another single metabolite output $m_2$. The stoichiometry of these networks is particularly simple: the columns $S^j$ of the stoichiometric matrix $S$ have at most one negative entry $-1$ and one positive entry $+1$.\\ 

%In particular, columns $S^{j^0_m}$ associated to outflow reactions $j^0_m$
%\begin{equation} \label{exit}
%j^0_m: \quad m \underset{j^0}{\longrightarrow} 
%\end{equation}
%have only a negative entry $-1$ in the $m^{th}$ row. \\

It is natural to model a monomolecular reaction network as a directed graph with a vertex metabolite set $\textbf{M}$ and an arrow reaction set $\textbf{E}$. We additionally require here that there are no self-loops, since a reaction $j$
\begin{equation}
j: \quad m \underset{j}{\longrightarrow} m,
\end{equation}
is without chemical meaning. A dipath (or directed path) is any acyclic ordered sequence of alternatingly adjacent vertices and arrows.\\

We state a simple characterization of the responses to a perturbation of a metabolite concentration $m^*$, in the monomolecular case. We have the following result:
\begin{thm} \label{monomet}
Let $m^*$ be a metabolite and $p'$ an element, either a metabolite $p'=m'$ or a reaction $p'=j'$, in a monomolecular reaction network. Then a perturbation of $m^*$ produces a response on an element $p'$ if and only if $p'$ is reachable from $m^*$ via a directed path.\\
\end{thm}
Above, the reachability is intended in the usual graph theory sense. That is, an element $p$ is reachable from a vertex $m$ if there exists a directed path $\gamma[m,p]$, which starts at the element $m$ and ends at the element $p$.\\

\section{Limitations to transitivity of influence}  \label{trans}

In previous works, great effort has been invested in the topic of transitivity of influence. Let us introduce transitivity by considering any $p_1$, $p_2$, and $p_3$ elements in the network, either metabolites or reactions. The \emph{transitivity question} is:
\begin{center}
\textit{If $p_1 \rightsquigarrow p_2$ and $p_2\rightsquigarrow p_3$, can we conclude that $p_1 \rightsquigarrow p_3$?}
\end{center}
The relevance of a positive answer to this question is both conceptual and practical. Conceptually, indeed, it explains the patterns observed experimentally in the responses. Practically, it greatly simplifies the computation of the nonzero responses.\\

The problem has been addressed in the literature for the reaction perturbation case, only. In the case of reaction perturbations, nonzero transitivity has been established in the monomolecular case \cite{FM15}, \cite{VM17}, at first. The general multimolecular case was resolved in \cite{BF18}. The result by Brehm and Fielder in \cite{BF18} actually claims more than the pure reaction perturbation case, and it is worth to recall it here:
\begin{thm}[Brehm-Fiedler] \label{BFtrans}
Let $p_1$ and $p_2$ be elements in a metabolic network. Let $j'$ be any reaction and $m'$ one of its input metabolites.
\begin{enumerate}
\item If $p_1 \rightsquigarrow m'$ and $j'\rightsquigarrow p_2$,  \quad \, then $p_1\rightsquigarrow p_2$. 
\item If $p_1\rightsquigarrow j'$ and $j' \rightsquigarrow p_2$, \quad \; then $p_1 \rightsquigarrow p_2$.
\end{enumerate}
\end{thm}
However the general case
\begin{center}
$p_1 \rightsquigarrow p_2$ and $p_2\rightsquigarrow p_3$  $\xRightarrow{?}$ $p_1 \rightsquigarrow p_3$
\end{center}
has remained open, for $p_2=m$ metabolite. \\

We show, with an extremely simple example, that Theorem \ref{BFtrans} cannot be improved, for the general multimolecular case. That is, any further transitivity claim of this type fails, in the multimolecular case. In the much simpler case of monomolecular networks, however, Theorem \ref{monotrans} extends transitivity to the pure metabolite case.\\ 

The simple example here is again Example \eqref{partialchild}:
\begin{equation} \label{countertrans}
\end{equation}

\vspace{-0.7cm}

\begin{center}
\includegraphics[scale=0.3]{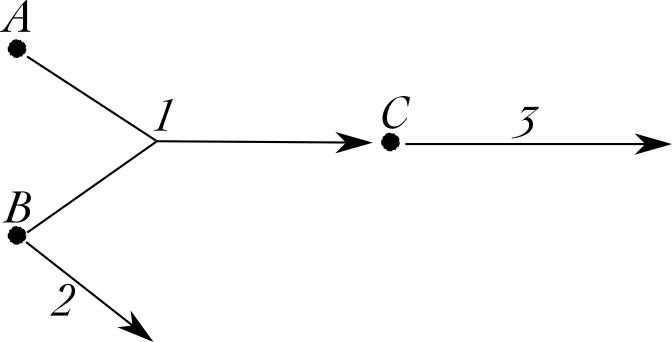}
\end{center}

In such small dimension, the full sensitivity matrix $\Psi$ can be computed explicitly. We omit the computation and we just report the sign of the entries of the matrix.
\begin{center}
$\Psi=$
\begin{tabular}{|c||c|c|c|c|c|c|}
\hline
$\downlsquigarrow$  & 1 & 2 & 3 & A & B & C \\ \hline \hline
1 & 0 & 0 & 0 & + & 0 & 0 \\ \hline
2 & 0 & 0 & 0 & - & + & 0 \\ \hline
3 & 0 & 0 & 0 & + & 0 & + \\ \hline
A & - & + & 0 & + & - & 0 \\ \hline
B & 0 & - & 0 & - & + & 0 \\ \hline
C & 0 & 0 & - & + & 0 & + \\ \hline
\end{tabular}
.
\end{center}

The four counterexamples are:
\begin{enumerate}[itemsep=0pt]
\item $B\rightsquigarrow A$ and $A\rightsquigarrow C$, but $B\not\rightsquigarrow C$;
\item $B\rightsquigarrow A$ and $A\rightsquigarrow 3$, but $B\not\rightsquigarrow 3$;
\item $2\rightsquigarrow A$ and $A\rightsquigarrow C$, but $2\not\rightsquigarrow C$;
\item $2\rightsquigarrow A$ and $A\rightsquigarrow 1$, but $2\not\rightsquigarrow 1$.
\end{enumerate}
Hence, Theorem \ref{BFtrans} covers all the transitivity properties.\\

In monomolecular reaction networks, however, the following theorem holds.
\begin{thm} [Monomolecular Transitivity] \label{monotrans}
Let $m^*$, $p_1$, $p_2$, be three elements of a monomolecular reaction network, where $m^*$ is a metabolite, and $p_1$ and $p_2$ are metabolites or reactions. Assume moreover that
\begin{center}
$m^*\rightsquigarrow p_1$ and $p_1\rightsquigarrow p_2$.
\end{center}
Then,
$$m^*\rightsquigarrow p_2.$$
\end{thm}

Even in the monomolecular case, nonetheless,
$$j \rightsquigarrow m \quad \quad \quad \text{and}  \quad \quad \quad m \rightsquigarrow p \quad \quad \quad \not{\Rightarrow}  \quad \quad \quad j \rightsquigarrow p.$$
Indeed, consider the following example.

\begin{equation}
\end{equation}

\vspace{-1cm}

\begin{center}
\includegraphics[scale=0.30]{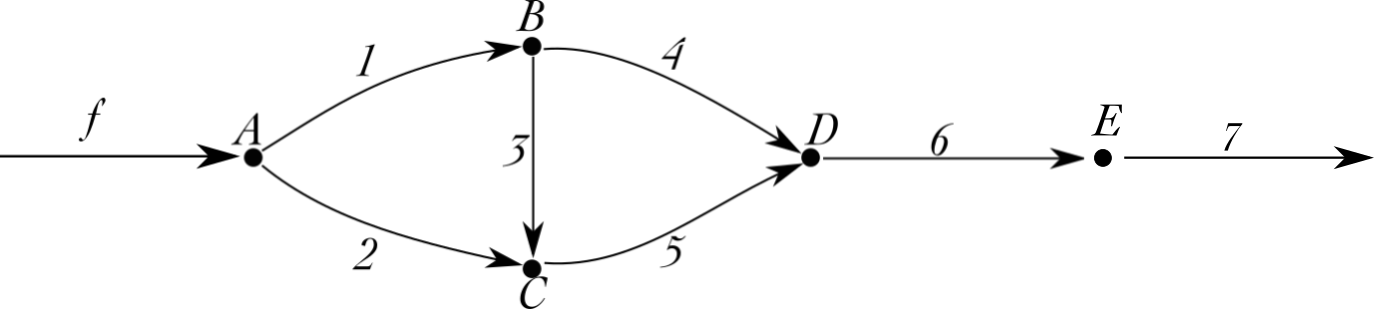}
\end{center}

Here, we know from \cite{FM15, VM17} that 

$$1 \rightsquigarrow B \quad \quad \quad \quad \text{but} \quad \quad \quad \quad 1 \not\rightsquigarrow 6 \quad \text{and} \quad  1 \not\rightsquigarrow E.$$

On the other hand, Theorem \ref{monomet}, shows that $B \rightsquigarrow 6$, $B \rightsquigarrow E$, since both the reaction arrow 6 and the metabolite vertex $E$ are reachable from the metabolite vertex $B$. In particular, then,

$$1 \rightsquigarrow B \rightsquigarrow 6\quad \text{and} \quad 1 \rightsquigarrow B \rightsquigarrow   E  \quad \quad \quad \quad \text{but} \quad \quad \quad \quad 1 \not\rightsquigarrow 6 \quad \text{and} \quad 1 \not\rightsquigarrow E.$$

\section{Discussion}\label{discussion}

In this paper, we have presented a sensitivity analysis for metabolic networks. We have focused on the case of \emph{perturbation of metabolite concentrations} and we have algebraically described the responses of the concentrations themselves and of the reaction fluxes. Here, \emph{algebraically} means that we have studied the responses as symbolic functions of the partial derivatives $r_{jm}$ of the reaction rates $\mathbf{r}(x)$ with respect to the concentrations $x$. Sufficiently parametrically rich kinetics, e.g. Michaelis-Menten, is best suited for our analysis, and it is usually anyway required to properly model a metabolic context, where enzymatic kinetics is often employed. In this case, we may conclude with no further analysis that there exist explicit reaction rates parameters for which an algebraically nonzero response corresponds factually to a nonzero response, evaluated at the equilibrium. On the contrary, the parametrically poor mass action case displays only one parameter for each reaction rate: for this reason, mass action would require further analysis for an analogous conclusion, within our approach.\\

We have provided a structural and qualitative analysis based on the stoichiometry of the network, only, and developed with the language of Child Selections, Theorems \ref{metmet} and \ref{fluxmet}. These results should be used as ground basis for further work, such as the sign analysis commented below. Moreover, there are also some immediate and counterintuitive consequences, which we have discussed in this paper. Firstly, self-influence does not always happen (Remark \ref{noself}). That is: even in the case of a direct external perturbation of a metabolite $m^*$, the new perturbed equilibrium may display the same unchanged concentration of $m^*$. Secondly, using the results of this paper jointly with the parallel results of \cite{BF18} about reaction perturbations, we have concluded that, mathematically, we can interpret a  perturbation of the metabolite $m^*$ simply as a reaction perturbation of an inflow reaction to $m^*$. Finally, transitivity of influence does not hold, in general, for perturbations of metabolites (Example \eqref{countertrans}).\\

Some delicate issues have not fully been addressed in detail in this paper.\\ 
Firstly, our approach is based on the Implicit Function Theorem. In this sense, our analysis is a \emph{local} analysis. However, in real applications, the fluctuation of parameters modeled by a perturbation may not be small. Fortunately, our results also apply to large perturbations, due to an interpolation argument by Brehm and Fiedler \cite{BF18}, which we omit here. Suffice it to say that the core reason, why such an interpolation argument holds, is that our analysis is only based on the stoichiometry of the system, and does not really depend - mathematically - on the specific equilibrium: the responses are algebraically the same, at any equilibrium.\\
Secondly, we have posed a nondegeneracy assumption \eqref{nd} throughout the paper, requiring the Jacobian determinant to be nonzero. This assumption is harmless on the intended application to metabolic networks. In fact, in a metabolic context this assumption practically always holds, due to the abundant presence of outflows. Without any desired application requiring a different context, we have restricted our presentation to this nondegenerate case. Mathematically possible, an extension of the present theory allowing stoichiometric subspaces has not been presented here, to avoid an overload of mathematical abstraction and confusing notation.\\

One natural next step in this sensitivity analysis is the study of the \emph{sign} of the responses. Monotonicity of the reaction rates, $r_{jm} > 0$, must be further assumed to make sense of this question. Then we can interpret the sign of \eqref{Bformulametmet} and \eqref{Bformulafluxmet} as rational functions of the positive variables $r_{jm}$. Thus, the rational function is \emph{positive, negative} or \emph{indeterminate}, depending whether it is positive for all values of the derivatives $r_{jm}$, negative, or the sign depends on those values. Preliminarily, a sign analysis of the Jacobian determinant $\operatorname{det}SR$, appearing as denominator of those formulas, is required. The sign of the Jacobian determinant, also in the context of zero eigenvalues bifurcations of equilibria, has been addressed in \cite{VGB20}. In particular, a Jacobian of indeterminate sign may indicate a simultaneous switch of the sign of the responses at the sign switch of the Jacobian, and it is therefore a particularly interesting point in itself. For the simpler monomolecular case a complete sign analysis for the flux influence has been established in \cite{V17}. The general case has been treated in the doctoral thesis \cite{Vas20}, from which a dedicated paper is in preparation.\\

The applications of our theory are multifold. For instance, we have presented in this paper a rational possibility to \emph{indirectly influence} certain network components by a direct perturbation on `distant' network elements. This consideration, best jointly with a sign analysis, opens a door to the \emph{control} of those metabolic elements, which are unreachable by direct perturbations. Furthermore, and in a different direction, the very structure of a network is unfortunately not always fully understood. Our analysis, grounded on the network structure only, offers in particular suggestions for \emph{network reconstruction}. To achieve this, further future work is needed to address the \emph{inverse} problem to the one presented here. That is: given only an abstract sensitivity matrix for metabolite concentrations, we could investigate which possible reactions (i.e., which possible network) are consistent with the given sensitivity matrix. This approach, based on experimentally obtained sensitivity matrices, may greatly help in understanding unknown network structures.

\section{Proofs} \label{pfs}

%This section is devoted to the proofs of the results in this chapter. 

We start this section with the proofs of Theorems \ref{metmet} and \ref{fluxmet}. These two proofs are both based on Cramer's rule and the Cauchy-Binet formula. They have an analogous structure: we invert the Jacobian determinant $SR$ using Cramer's rule, obtaining:
\begin{equation}
((SR)^{-1})^{m^*}_{m'}=\frac{(-1)^{m^*+m'} \operatorname{det}((SR)^{\vee {m'}}_{{\vee m^*}})}{\operatorname{det}SR},
\end{equation}
where $(SR)^{\vee {m'}}_{{\vee m^*}}$ indicates the submatrix of $SR$ taken removing the $m^{*\;th}$ row and the $m'^{\;th}$ column. The numerator is analyzed via Cauchy-Binet formula and interpreted in terms of Child Selections, according to each theorem. One word on notation: for a matrix $\mathpzc{A}$, the notation $\mathpzc{A}^\mathcal{E}_\mathcal{F}$ denotes the submatrix of $\mathpzc{A}$ consisting of columns $\mathcal{E}$ and rows $\mathcal{F}$. For simplicity of notation, we have omitted the braces $\{m\}$ for single elements, so that, for example, $R_j$ indicates the $j^{th}$ row and $R^m$ indicates the $m^{th}$ column of $R$. We list the proofs one after the other.\\

\proof [Proof of Theorem \ref{metmet}]
The response $(\delta x)^{m^*}_{m'}$ of a metabolite concentration $m'$ to a perturbation of the metabolite concentration $m^*$ is given by
\begin{equation}
(\delta x)^{m^*}_{m'}=-((SR)^{-1})^{m^*}_{m'},
\end{equation}
where $((SR)^{-1})^{m^*}_{m'}$ indicates the entry corresponding to the $m^{*\;th}$ column and $m'^{\;th}$ row. We invert the Jacobian $SR$ using Cramer's rule:
\begin{equation}
\begin{split}
(\delta x)^{m^*}_{m'}=-((SR)^{-1})^{m^*}_{m'}=-\frac{(-1)^{m^*+m'} \operatorname{det}((SR)^{\vee {m'}}_{{\vee m^*}})}{\operatorname{det}SR}.
\end{split}
\end{equation}
Here, again, $(SR)^{\vee {m'}}_{{\vee m^*}}$ indicates the submatrix of $SR$ taken removing the $m^{*\;th}$ row and the column $m'^{\;th}$ column. We analyze the numerator, using Cauchy-Binet formula:
\begin{equation}
\begin{split}
- \operatorname{det}SR \, (\delta x)^{m^*}_{m'}=& (-1)^{m^*+m'} \operatorname{det}((SR)^{\vee {m'}}_{{\vee m^*}})\\ 
=&(-1)^{m^*+m'} \operatorname{det}(S_{\vee {m^*}}R^{{\vee m'}})\\
=&{(-1)^{m^* + m'} \sum \limits_{\mathcal{E}' \in \mathcal{E}^{M-1}} \operatorname{det} S^{\mathcal{E}'}_{\vee m^*} \cdot \operatorname{det} R_{\mathcal{E}'}^{\vee m'}}.
\end{split} 
\end{equation}

We observe that $\operatorname{det}R_{\mathcal{E}'}^{\vee m'}\neq 0$ if and only if there exists a Partial Child Selection $\mathbf{J^{\vee m'}}$: $\mathbf{M} \smallsetminus \{ m' \} \longrightarrow \mathbf{E}$, such that $\mathbf{J^{\vee m'}}(\mathbf{M} \smallsetminus \{ m' \})= \mathcal{E}'$. This observation and the signature argument
\begin{equation}
\operatorname{sgn}(\mathbf{J^{\vee m'}}) \, \operatorname{det}S^{ \mathcal{E}' =\mathbf{J^{\vee m'}}(\mathbf{M} \smallsetminus m')}=  \operatorname{det} S^{{\mathbf{J^{\vee m'}}}}_{\vee m^*}, 
\end{equation}
leads to the equality: 
\begin{equation} 
(\delta x)^{m^*}_{m'}=-\frac{(-1)^{m^*+m'} \sum \limits_{\mathbf{J^{\vee m'}}} \operatorname{det} S^{{\mathbf{J^{\vee m'}}}}_{\vee m^*} \cdot \prod_{m\in \mathbf{\mathbf{M}}\setminus {m'}} r_{({\mathbf{J^{\vee m'}}(m))m}}}{\operatorname{det}SR},
\end{equation}
and the following last equality, due to Laplace determinant expansion, concludes the proof.
\begin{equation}
(-1)^{m^*+m'}\operatorname{det}S_{\vee m^*}^{\mathbf{J^{\vee m'}}}=\operatorname{det}S^{\mathbf{J^{\vee m'}\cup e_{m^*}}}.
\end{equation}
\endproof

\proof[Proof of Theorem \ref{fluxmet}]
Analogously as above, we analyze the expression  $(\Phi)^{m^*}_{j'}=-R_{j'} ((SR)^{-1})^{m^*}$ for the flux response of $j'$ to a metabolite perturbation of $m^*$.
\begin{equation}
\begin{split}
(\Phi)^{m^*}_{j'}=&- R_{j'} ((SR)^{-1})^{m^*}\\
=& - \sum_{m\in{\mathbf{M}}} R_{j'}^{m} ((SR)^{-1})^{m^*}_{m}\\
=&- \sum_{m\in{\mathbf{M}}} R_{j'}^{m} \frac{(-1)^{m^*+m} \operatorname{det}(SR)^{\vee {m}}_{{\vee m^*}}}{\operatorname{det}SR}.
\end{split}
\end{equation}
That is,
\begin{equation}
\begin{split}
-\operatorname{det}SR \, (\Phi)^{m^*}_{j'}=& \sum_{m\in{\mathbf{M}}} R_{j'}^{m} (-1)^{m^*+m} \operatorname{det}(SR)^{\vee {m}}_{{\vee m^*}}\\
=& \sum_{m\in{\mathbf{M}}} R_{j'}^{m} {(-1)^{m^*+ m} \sum \limits_{\mathcal{E}' \in \mathcal{E}^{M-1}} \operatorname{det} S^{\mathcal{E}'}_{\vee m^*} \cdot \operatorname{det} R_{\mathcal{E}'}^{\vee m}}\\
=& \sum \limits_{\mathcal{E}' \in \mathcal{E}^{M-1}} \sum_{m\in{\mathbf{M}}} ((-1)^{m^*+ j'} \operatorname{det} S^{\mathcal{E}'}_{\vee m^*}) ((-1)^{m+ j'} R_{j'}^{m} \operatorname{det} R_{\mathcal{E}'}^{\vee m})\\
=&\sum \limits_{\mathcal{E}' \in \mathcal{E}^{M-1}} \operatorname{det} S^{\mathcal{E}'\cup e_{m^*}} \operatorname{det}R_{\mathcal{E}'\cup j'}\\
=& \sum \limits_{\mathbf{J}\ni j'}  \operatorname{det} S^{\mathbf{J(M)}\smallsetminus  {j'} \cup e_{m^*}} \operatorname{sgn}(\mathbf{J}) \prod_{m\in \mathbf{M}} r_{\mathbf{J}(m)m}\\
=&  \sum \limits_{\mathbf{J}\ni j'} \operatorname{det} S^{\mathbf{J} \smallsetminus  {j'} \cup e_{m^*}} \prod_{m\in \mathbf{M}} r_{\mathbf{J}(m)m}.
\end{split}
\end{equation}

\endproof

 %To be clear, in the present work, we present the corollary as a corollary to the independently proven results ... and \cite{BF18}. Of course, another possibility would be to prove independently - as a theorem, the corollary and read then ...

The proof of Corollary \ref{corred} requires a technical recall on formulas from \cite{BF18}. In particular, parallel formulas to \eqref{Bformulametmet} (metabolite response) and  \eqref{Bformulafluxmet} (reaction flux response) were derived in the case of reaction perturbation. The formula for the response $(\delta x)^{j^*}_{m'}$ of the metabolite $m'$ to a perturbation of the reaction $j^*$ reads:
%\begin{equation*}
%\delta x)^{m^*}_{m'}=-\frac{(-1)^{m^*+m'} \sum \limits_{\mathbf{J^{\vee m'}}} \operatorname{det} S^{{\mathbf{J^{\vee m'}}}}_{\vee m^*} \cdot \prod_{m\in \mathbf{M}\smallsetminus {m'}} r_{{\mathbf{J^{\vee m'}}(m)m}}}{\operatorname{det}SR}.
%\end{equation*}
%and for $(\Phi)^{m^*}_{j'}$, 
%\begin{equation*}
%(\Phi)^{m^*}_{j'}=-\frac{ \sum \limits_{\mathbf{J}\ni j} \operatorname{det}S^{\mathbf{J} \setminus  j \cup {e_m^*}}  \prod_{m\in \mathbf{M}} r_{\mathbf{J}(m),m}}{\operatorname{det}SR},
%\end{equation*}
%have been derived in the language of Child Selections. The formula for the metabolite response to reaction perturbation reads:
\begin{equation} \label{fbmetmet}
(\delta x)^{j^*}_{m'}= - \frac{\sum_{\mathbf{J^{\vee {m'}}}\not \ni {j^*}} \operatorname{det}S^{\mathbf{J}^{\mathbf{\vee^{m'}} \cup {j^*}}} \cdot \prod_{m \in \mathbf{M} \setminus m'} r_{\mathbf{J^{\vee {m'}}}(m)m}}{\operatorname{det}SR},
\end{equation}
and the formula for the response $(\Phi)^{j^*}_{j'}$ of the flux of $j'$ to a perturbation of $j^* \neq j'$ reads
\begin{equation}\label{fbfluxmet}
(\Phi)^{j^*}_{j'}=-\frac{\sum_{{j^*}\not\in \mathbf{J} \ni {j'}}\operatorname{det}S^{\mathbf{J} \smallsetminus {j'} \cup {j^*}}\prod_{m\in \mathbf{M}}r_{\mathbf{J}(m)m}}{\operatorname{det}SR}.
\end{equation}

We are now ready for the proof of Corollary \ref{corred}.
\proof[Proof of Corollary \ref{corred}]
The proof consists only in noticing that, for a inflow reaction $f_{m^*}$ to $m^*$, the stoichiometric column $f_{m^*}$ \emph{is} the unit vector $e_{m^*} \in \mathbb{R}^M$, establishing equivalency of the formulas \eqref{fbmetmet} with \eqref{Bformulametmet} (and \eqref{fbfluxmet} with \eqref{Bformulafluxmet}, respectively).
%For the singular self-influence case $(\Phi)^{f_{m^*}}_{f_{m^*}}$ just note that 
%$$(\Phi)^{f_{m^*}}_{f_{m^*}}=1- R_{f_{m^*}} (SR)^-1 S^{f_{m^*}}=1-(\Phi)^{m^*}_{f_{m^*}},$$
%where the last equality simply follows from the previous consideration on the formulas.
\endproof

Next, we prove Theorem \ref{monomet} about monomolecular networks. Again, a small technical device is required, to properly represent monomolecular networks as directed graph; we introduce a further artificial vertex: the \emph{zero-complex} $\mathbf{0}$. The zero-complex \textbf{0} in the words of Feinberg \cite{Fei87} is \emph{`a complex in which the stoichiometric coefficient of every species is zero'}. In the multimolecular context of the previous chapters, we have avoided introducing the zero-complex, as a superfluous tool. On the contrary, it is of use in the monomolecular setting. In particular, it is required to model inflow and outflow reactions. An outflow reaction from metabolite $m$ is represented with the zero-complex as
\begin{equation} 
j^0_m: \quad m \underset{j^0_m}{\longrightarrow} \mathbf{0}.
\end{equation}
An inflow reaction to metabolite $m$ is represented with the zero complex as
\begin{equation}
f^0_m: \quad \mathbf{0} \underset{f^0_m}{\longrightarrow} m.
\end{equation}
In \cite{FM15}, Fiedler and Mochizuki proved a theorem for the responses to reaction perturbations in monomolecular networks, characterizing the nonzero responses only via graph means. 
\begin{thm} [Fiedler\&Mochizuki] \label{FM}
Consider any pair of ($j^*$, $m'$), where $j^*$ is a reaction and $m'$ a metabolite. Then the response $(\delta x)^{j^*}_{m'}$ of $m'$ to a perturbation of $j^*$ is nonzero, algebraically, if and only if there exist two dipaths $\gamma^0$ and $\gamma'$ such that:
\begin{enumerate}[itemsep=0pt]
\item both dipaths emanate from $m^*$, input metabolite of $j^*$;
\item one of the dipaths contains $j^*$;
\item $\gamma^0$ terminates at vertex $\textbf{0}$, and $\gamma'$ terminates with the metabolite $m'$;
\item except for their shared starting vertex $m^*$, the two dipaths $\gamma^0$ and $\gamma'$ are disjoint.
\end{enumerate}
Moreover, for a reaction $j' \neq j^*$, such that $m'$ is an input to $j'$, we have
$$(\Phi)^{j^*}_{j'} \neq 0,\quad \text{ algebraically} \quad \Leftrightarrow \quad (\delta x)^{j^*}_{m'} \quad  \text{ algebraically}.$$
\end{thm}
Now we concentrate on the proof of Theorem \ref{monomet}.
\proof[Proof of Theorem \ref{monomet}]
According to Corollary \ref{corred}, a metabolite perturbation of $m^*$ is identical to a reaction perturbation of an inflow reaction $f^0_{m^*}$ to $m^*$. Consider now Theorem \ref{FM}: a nonzero response of an element $p'$ (either metabolite or reaction) to a perturbation of $m^*$ is equivalent to the existence of two dipaths $(\gamma^0, \gamma')$, both departing from $\mathbf{0}$, `mother' of $f^0_{m^*}$: $\gamma^0$ leading to $0$, $\gamma'$ leading to $p'$. In our case, we should choose as $\gamma^0$ the trivial path $\gamma^0\equiv \{ \mathbf{0}\}$. This always provides the existence of $\gamma^0$ satisfying the needed conditions. For any reachable $p'$ from $m^*$, any directed path $\gamma[0,p']$, starting with the arrow $f^{0}_{m^*}$, serves as $\gamma'$, providing the desired pair $(\gamma^0, \gamma')$.\\
\endproof

Remaining with monomolecular networks, we conclude this section with the proof of Theorem \ref{monotrans}.

\proof[Proof of Theorem \ref{monotrans}]
The case in which $p_1$ is a reaction reduces to Theorem \ref{BFtrans}. Next, we only need to consider $p_1=m$ metabolite. This case is solved by the monomolecular Theorem \ref{monomet} for metabolite influence, which states that metabolite influence is equivalent to reachability. Reachability in networks is obviously a transitive property.
\endproof

\bibliography{bibliography/references.bib}
 \bibliographystyle{ieeetr}

\end{document}